\documentclass[12pt]{article}
\usepackage{amsmath}
\usepackage{latexsym}
\usepackage{a4wide}
\usepackage{bbm}
\usepackage{epsfig}
\usepackage{feynmf}
\newcommand{\I}{\text{i}}
\newcommand{\E}{\text{e}}
\newcommand{\tr}{\text{tr}}

\newcommand{\sta}[1]{{}^\star\! #1}
\newcommand{\re}[1]{(\ref{#1})}
\newcommand{\GF}{G_{\text{F}}}
\newcommand{\gV}{g_{\text{V}}}
\newcommand{\gA}{g_{\text{A}}}
\newcommand{\tW}{\theta_{\text{W}}}
\newcommand{\Geff}{\Gamma_{\text{eff}}}
\newcommand{\Bcr}{B_{\text{cr}}}
\begin{document}
\abovedisplayskip14pt plus2pt minus4pt
\abovedisplayshortskip7pt plus2pt minus4pt
\belowdisplayskip14pt plus2pt minus4pt
\belowdisplayshortskip7pt plus2pt minus4pt
\title{\Large \bf Neutrino interactions with a weak slowly varying
  electromagnetic field \thanks{Published in Phys.\ Lett.\  {\bf
      B480}, 129 (2000) \copyright 2000 Elsevier Science B.V.}}
\author{\large Holger Gies and Rashid Shaisultanov\\
  {\small \it Institut f\"ur theoretische Physik, Universit\"at
    T\"ubingen,}\\ 
  {\small \it 72076 T\"ubingen, Germany}}
\maketitle
\begin{abstract}
We derive the effective action for processes involving two neutrinos
and two photons at energies much below the electron mass. We discuss
several applications in which one or both photons are replaced by
external fields. In particular, Cherenkov radiation and neutrino pair
production in weak external fields are investigated for massive Dirac
neutrinos.  
\end{abstract}

\section{Introduction}

The subject of electromagnetic interactions of neutrinos has been
widely discussed in the literature, due to their relevance to
astrophysics and cosmology. 
In the standard model, neutrino-photon interactions appear at the
one-loop level. To be precise, it is the charged particle running in
the loop which, when integrated out, confers its electromagnetic
properties to the neutrino. This induces an effective coupling between
photons and neutrinos. 
E.g., a process involving two neutrino legs and one photon (e.g.,
$\nu\to\nu\gamma$) which is forbidden in vacuum can become important
in the presence of a medium and/or external fields \cite{doli89,milt76}.
The standard model processes with two photons, for example,
neutrino-photon scattering \( \gamma \, \, \nu \rightarrow \gamma \,
\, \nu \, \, \), turn out to be highly suppressed in vacuum. In
\cite{gell61} Gell-Mann showed that in the four-Fermi limit of the
standard model the amplitude is exactly zero to order $\GF$; this is
because, according to Yang's theorem \cite{yang50}, which is based on
rotational invariance, two photons cannot couple to a \( J=1 \) state.
Therefore the amplitude is suppressed by the additional factors of \(
\omega /m_{W} \), where \( \omega \) is the photon energy and \( m_{W}
\) is the \( W \) mass \cite{gell61,a2,b,c}.  As a result, the typical
vacuum cross sections are exceedingly small (e.g., see \cite{f} for
the process $\gamma\nu\to\gamma\nu$).

The same reactions in the presence of an external magnetic field are
enhanced by the factor \( \sim \left( m_{W}\, \, /m_{e}\right)
^{4}\left( B/B_{c}\right) ^{2} \) for $ \omega \ll m_{e}$ and $B \ll
B_{c}$ as was shown in \cite{shai98} (extensions to this result can be
found in \cite{vas,kit}).

In the present letter, we investigate the interaction of two neutrinos
with an electromagnetic field to lowest, i.e., second, order in the
field by deriving the corresponding effective action. The field can be
considered as slowly varying (compared to the Compton wavelength) or,
alternatively, as produced by two soft off-shell photons. Since such a
field can have any state of angular momentum, the amplitude is
generally not suppressed, due to Yang's theorem. A second possibility
to circumvent Yang's theorem arises from non-vanishing Dirac neutrino
masses, which allow for a $\bar{\nu}\nu$-pair in a $J=0$ state.

The calculation as outlined in the following section is actually very
simple and, within the four-Fermi limit of the electroweak theory, can
solely be based on the famous triangle diagram. The resulting
effective action easily reproduces well-known results and finally
reveals a variety of interesting new effects that will be briefly
discussed in Sec. \ref{Seceffects}; it includes the production of
massive as well as massless neutrinos by varying electromagnetic
fields, and Cherenkov radiation for massive neutrinos. We furthermore
hint at an enhancement of neutrino oscillations by varying fields
without an additional medium.

\section{Effective $\nu\nu\gamma\gamma$ Action}

For the derivation of the desired effective action, we employ a
set of approximations: first, we assume the neutrino energies to be
very much smaller than the $W$- and $Z$-boson masses, allowing us to
use the local limit, i.e., the effective four-Fermi interaction:
\begin{equation}
{\cal L}=\frac{\GF}{\sqrt{2}}\, \bar{\nu}\gamma_\mu (1+\gamma_5)
\nu\, \bar{E}\gamma^\mu (\gV +\gA \gamma_5) E. \label{1}
\end{equation}
Here, $E$ denotes the electron field, $\gV=\frac{1}{2} +2\sin^2\tW$
and $\gA=\frac{1}{2}$ for $\nu_e$, and $\gV=-\frac{1}{2} +2\sin^2 \tW$
and $\gA=-\frac{1}{2}$ for $\nu_{\mu,\tau}$. 

Now we place the system into an external electromagnetic field to
which the charged fermions can couple, implying the effective action
%%%%%%%%%%%%%%%%%%%%%%%%%%%%%%%%%%%%%%%%%%%%%
% 10.03.00
% misprint: forgotten spacetime integral
%%%%%%%%%%%%%%%%%%%%%%%%%%%%%%%%%%%%%%%%%%%%%
\begin{equation}
\Geff[L,A]= \frac{\GF}{\sqrt{2}} \frac{1}{e}\int d^4x\, L_\mu \bigl(
\gV \langle j^\mu\rangle^A +\gA \langle j_5^\mu\rangle^A \bigr),
\label{2} 
\end{equation}
where we introduced the neutrino current $L_\mu:=\bar{\nu}\gamma_\mu
(1+\gamma_5) \nu$ and the electromagnetic currents $j_5^\mu:=
e \bar{E} \gamma^\mu \gamma_5 E$ and $ j^\mu:=
e \bar{E} \gamma^\mu E$; their expectation values in an
external field can be reexpressed in terms of the Green's function $G$
in this field:
\begin{equation}
\langle j_5^\mu\rangle^A= \I e\tr\bigl[  \gamma^\mu \gamma_5\, G(x,x|A)
\bigr], \label{3}
\end{equation}
and similarly for $j^\mu$. Obviously, the current expectation values
correspond to closed fermion loops coupled to the external field. In
the following, we omit all contributions from any charged loop fermion
other than electron-positron pairs, since they are suppressed by
inverse powers of their mass. Due to the properties of the Dirac
algebra (Furry's theorem), only odd numbers of external field
couplings contribute to the vector current $\langle j^\mu\rangle^A$,
while only even numbers contribute to the axial one $\langle
j_5^\mu\rangle^A$.

Since we are furthermore interested in electromagnetic fields whose
strength and variation is bound by the scale of the electron mass $m$,
the lowest-order non-trivial contribution to $\Geff$ in a weak-field
expansion arises from the axial current. Expanding the axial current
to second order in $A^\mu$ leads us to
\begin{eqnarray}
\langle j_{5\lambda} (x)\rangle\!\!\!\! &=&\!\!\!\!
 \frac{\I e}{2}\!\! \int\!\! \frac{d^4k_1\,
  d^4k_2\, d^4k}{(2\pi)^{12}}  \E^{\I x(k_1+k_2)}\, eA^\mu(k_1)\,
  eA^\nu(k_2)\, \tr\, \Bigl[ g(k_1+k) \gamma_\mu g(k) \gamma_\nu
  g(k-k_2) \gamma_\lambda \gamma_5 \Bigr]\nonumber\\
&& + \Bigl\{ \mu
  \leftrightarrow \nu\Bigr\} \nonumber\\
&=:& \frac{e}{2} \int \frac{d^4k_1\, d^4k_2}{(2\pi)^8}\, 
\E^{\I x(k_1+k_2)}\, eA^\mu(k_1)\,  eA^\nu(k_2)\,
  \Delta^5_{\mu\nu\lambda}(k_1,k_2). \label{4}
\end{eqnarray}
Here we employed the Fourier representations of the free Green's
function $g(p)$ as well as the external field $A(p)$. In the last line
of Eq. \re{4}, we identified the gauge invariant amplitude of the
famous triangle diagram, where $k_1$ and $k_2$ are the photons'
momenta and $k$ runs around the loop. We may borrow the final result
for $\Delta^5_{\mu\nu\lambda}$ from, e.g., \cite{crew82}. Since we are
working in the soft-photon approximation, i.e., $k_1,k_2\ll m$, the
amplitude simplifies considerably and reduces to a sum of terms which
are linear in $k_1$ and quadratic in $k_2$ or vice versa. After
Fourier transforming them back into coordinate space, we obtain the
lowest order term of the axial current in a weak-field and
soft-momentum expansion\footnote{In principle, the axial current in
  Eq. \re{3} is an ill-defined operator identity because of the
  presence of the anomaly; however, in the here-considered ``large
  electron mass'' expansion, the anomaly is not present, since it is
  mass independent. In other words, the anomaly does not contribute to
  the neutrino-photon interaction as studied in this work.}:
\begin{equation}
\langle j_5^\mu (x)\rangle=\frac{\alpha}{6\pi} \, \frac{e}{m^2}
\Bigl( \partial^\mu {\cal G} + (\partial^\alpha F_{\alpha\beta})
\sta{F}^{\beta\mu} \Bigr) +{\cal O}(1/m^6), \label{5}
\end{equation}
where we employed the conventions $\sta{F}^{\alpha\beta}=\frac{1}{2}
\epsilon^{\alpha\beta\kappa\lambda} F_{\kappa\lambda}$, and ${\cal
  G}:=-\frac{1}{4} F^{\mu\nu} \sta{F}_{\mu\nu}$. Upon insertion into
Eq. \re{2}, the effective action for the lowest-order neutrino-photon
interaction reads: 
\begin{equation}
\Geff[L,A]=\frac{\GF \gA}{\sqrt{2}} \frac{\alpha}{6\pi}
\frac{1}{m^2}\, \int d^4x \Bigl( -(\partial^\mu L_\mu)\, {\cal G} +
(\partial^\alpha F_{\alpha\beta})\, (L_\mu \sta{F}^{\beta\mu}) \Bigr)
+{\cal O}(1/m^4). \label{6}
\end{equation}
This equation represents the central result of our investigation. Now,
if all external particles are on shell, we get $\partial_\mu L^\mu=0$
for massless neutrinos, and $\partial_\alpha F^{\alpha\beta}=0$ for
free photons. Therefore, the amplitude vanishes, which is nothing but
the manifestation of Yang's theorem. However, interesting effects can
be discovered for deviations from this (standard model) on-shell
behavior. A first glance at this will be outlined in the next
section. 

Let us finally mention that Eq. \re{6} represents the second-order
analogue of the effective action derived by Dicus and Repko
\cite{dicu97}, which describes neutrino interactions with three
external photon lines; the latter characterizes the lowest-order
effective theory for (standard model) on-shell interactions.

\section{Applications}
\label{Seceffects}
By construction, Eq. \re{6} represents the effective action for
two-photon two-neutrino processes, and we can easily derive the matrix
element for, e.g., the process $\bar{\nu}\nu\to\gamma\gamma$ for
massive neutrinos:
\begin{equation}
{\cal M}(\bar{\nu}_l \nu_l\to \gamma(k,\epsilon),
\gamma(k',\epsilon')) =-\frac{\GF (2\gA)}{\sqrt{2}}
\frac{\alpha}{6\pi} \frac{m_\nu}{m^2} \, \bar{v}^{\bar{\nu}}_l
\gamma_5 u^\nu_l\, 
\epsilon^{\mu\nu\alpha\beta} k_\mu k'_\nu \epsilon_\alpha
\epsilon'_\beta, \label{7}
\end{equation}
which is in perfect agreement with \cite{crew82} for $\gA=\frac{1}{2}$
for $\nu_e$ and $\gA=-\frac{1}{2}$ for $\nu_{\mu,\tau}$, as it should
be.  We observe that Eq. \re{7} arises from the first term of Eq.
\re{6} only, since the second term vanishes for on-shell photons.

\subsection{Cherenkov radiation by massive neutrinos in magnetic
  fields} 

A review of this effect for massless neutrinos has been performed in
study \cite{ioan97}. The emission of Cherenkov radiation by neutrinos
propagating perpendicular to a magnetic field becomes possible,
because the phase velocity of soft photons in a magnetic field is
smaller than in vacuum $k^2\sim - \alpha\frac{B^2}{\Bcr^2}$,
$\Bcr=\frac{m^2}{e}$ (for a review, see \cite{gies98}). It can be
shown that the contributions of the vector part of $\Geff$ in Eq.
\re{2} to the transition rate are proportional to $k^2$ and hence are
suppressed by additional orders of $\alpha^2$. The same holds for the
axial part of Eq. \re{2} if one considers $\bot$-photons which are
perpendicularly polarized compared to the field direction. Hence, only
parallelly polarized $\|$-photons can be emitted as Cherenkov
radiation.

Since $\partial^\alpha F_{\alpha\beta}\sim k^2 \epsilon_\beta$, the
contributions of the second-order effective action \re{6},
can similarly be neglected for massless neutrinos, so that the lowest
order contribution to the well-known transition rate arises from the
terms $\sim 1/m^6$ in $\Geff$ corresponding to a pentagon diagram. The
situation changes for massive neutrinos when the first term of
Eq. \re{6} no longer vanishes. 

Associating one field strength tensor with the external magnetic field
and the other with the emitted $\|$-photon of frequency $\omega$, the
matrix element reads:
\begin{equation}
{\cal M}(\nu(p)\to \nu(p'),\gamma(k))=\I \frac{\GF \gA}{\sqrt{2}}
\frac{\alpha}{3\pi} {m_\nu}\,\omega \frac{B}{m^2}\,
\bar{u}^\nu_l(p')\gamma_5 u^\nu_l(p). \label{8}
\end{equation}
For neutrino energies below the $e^+e^-$ pair production threshold,
and neglecting the small deviations from the photon light cone,
the transition rate yields:
\begin{eqnarray}
\Gamma_{\nu\to\nu\gamma}&=&\frac{1}{2E} \sum_{\text{pol.}}
\int\frac{d^3k}{(2\pi)^3 2\omega} \int \frac{d^3p'}{(2\pi)^3 2E'} \,
(2\pi)^4 \, \delta^4(p-(p'+k))\, |{\cal M}|^2 \nonumber\\
&=& \frac{7}{2^9\cdot3^4\cdot 5^2} \frac{\alpha^2}{\pi^4} \, m \, (\GF m^2)^2\,
\left(\frac{E}{m}\right)^3\left( 1-\frac{E_{\text{min}}^2}{E^2}
\right)^5 \left( \frac{m_\nu}{m}\right)^2 \left( 
  \frac{B}{\Bcr} \right)^4 \theta(E-E_{\text{min}})\nonumber\\
&=&2.6\cdot10^{-14}\,\text{s}^{-1}\, 
\left(\frac{E}{m}\right)^3\left( 1-\frac{E_{\text{min}}^2}{E^2}
\right)^5 \left( \frac{m_\nu}{m}\right)^2 \left( 
  \frac{B}{\Bcr} \right)^4 \theta(E-E_{\text{min}})\label{9}
\end{eqnarray}
where $\Bcr=\frac{m^2}{e}$ and 
\begin{equation}
E_{\text{min}}:= \sqrt{\frac{45\pi}{7\alpha}} m_\nu
\frac{\Bcr}{B}. \label{10}
\end{equation}
The existence of such a mass-dependent threshold energy
$E_{\text{min}}$ for the incoming neutrino arises from the Cherenkov
condition: the neutrino must move ``faster than light'' in the
$B$-field background. It is easy to check that the transition rate
\re{9} never wins out over the mass-independent contribution as cited
in \cite{ioan97}, which is proportional to $(E/m)^5(B/\Bcr)^6$. This
statement holds for any value of the neutrino mass in the low-energy
domain. 

We would like to stress that this zero-result is a non-trivial
statement and should serve as a counterexample for the general belief
that non-zero neutrino masses always open a window in parameter space
for new effects.  

\subsection{$\bar{\nu}\nu$-pair emission by varying electromagnetic
  fields} 

Although electron-positron pair emission by varying electromagnetic
fields belongs to standard textbook knowledge (see, e.g.,
\cite{itzy80}), it is far from being phenomenologically important, due
to the enormous threshold frequency $\omega_{\text{cr}}= 2 m=1.5\cdot
10^{21}$Hz with which the field must oscillate. The effective action
Eq. \re{6} reveals a similar mechanism for neutrinos which benefits
from the smallness of the neutrino mass.

Let us consider spacetime regions where a varying electromagnetic
field satisfies the vacuum Maxwell equations; then only the first term
of Eq. \re{6} contributes to the pair production matrix element:
\begin{equation}
{\cal M}({\cal G}(k) \to \bar{\nu}(p'),\nu(p))= \I
\frac{\GF\gA}{\sqrt{2}} \frac{\alpha}{3\pi} \frac{m_\nu}{m^2}\, {\cal
  G}(k)\, \bar{v}^{\bar{\nu}}_l(p') \gamma_5 u^\nu_l(p), \qquad
k=p+p', \label{11}
\end{equation}
where ${\cal G}(k)$ denotes the Fourier transform of ${\cal G}(x)$. The
production probability is given by
\begin{eqnarray}
W&=& \int d^4k |{\cal G}(k)|^2 \int \frac{d^3p}{(2\pi)^3 2E} \int
\frac{d^3p'}{(2\pi)^3 2E'} \, \delta^4(k-(p+p')) \sum_{\text{pol.}}
|{\cal M}|^2 \nonumber\\
&=&\frac{\GF^2 \alpha^2}{36 (2\pi)^7} \frac{m^2_\nu}{m^4} \int d^4k \,
|{\cal G}(k)|^2\, k^2\left( 1- \frac{4m_\nu^2}{k^2} \right)^{1/2}
\theta\left(1- \frac{4m_\nu^2}{k^2} \right). \label{12}
\end{eqnarray}
In order to obtain an illustrative estimate of the order of magnitude
of this effect, let us simply take ${\cal G}(k)= \mathbf{E_0\cdot B_0}
(2\pi)^4 \delta^3(\mathbf{k}) \delta(\omega -2\omega_0)$. For this
field configuration, we obtain the production probability per volume
and time
\begin{equation}
\frac{W}{V T} \simeq 5.11
 \left( \frac{m_\nu}{1\text{eV}}\right)^2
 \left(\frac{\omega_0}{1\text{eV}} \right)^2 \frac{(\mathbf{E_0\cdot
 B_0})^2}{\Bcr^4} \left(1-\frac{m_\nu^2}{\omega_0^2}\right)^{1/2}
 \theta (\omega_0-m_\nu)\, {\text{cm}^3}{\text{s}}, \label{13}
\end{equation}
in units of cm${}^3$ and seconds.  Obviously, the threshold frequency
is equal to the neutrino mass, e.g., in the strong ultraviolet for
neutrino masses at the eV-scale.  Equation \re{13} can also be
interpreted as the number of pairs produced in the system and volume
under consideration \cite{niki70}.

It is instructive to compare this pair-production
probability with the one for $e^+e^-$-pairs: $W_{e^+e^-}\sim \int
(E^2-B^2)$. Each process is triggered by a different invariant of the
electromagnetic field revealing its vector or axial vector character. 

Another neutrino pair-creation mechanism was proposed by Kachelriess
\cite{kach98}, which is caused by a density gradient of background
fermions (e.g., neutrons). It is interesting to observe that the two
mechanisms exhibit very different sensitivities to the neutrino mass:
while the above-studied mechanism obeys a power law, the latter decays
exponentially with increasing $m_\nu$, since this effect is based on
the Schwinger mechanism.

\section{Comments}
In the present letter, we derived the low-energy effective action for
processes involving two neutrinos and two photons (or couplings to an
external field) and discussed several immediate applications.

The here-considered effects arise essentially from the first term of
the effective action in Eq. \re{6}. Of course, we could have also
studied field configurations inducing similar effects that are
triggered solely by the second term of Eq. \re{6}, e.g., purely
magnetic field configurations with spatial variation. In this case, we
should, however, confine the investigations to situations where the
fields are not produced by standard model currents, since those
currents will also couple directly and without ${\cal
  O}(\alpha)$-suppression to the neutrino current. Such fields might
be produced by, e.g., aligned spins, as is possible in neutron stars,
or variable Higgs condensates which might be responsible for
primordial magnetic fields.  Upon insertion of these field
configurations into the above-given calculations, the neutrino mass
terms get effectively replaced by spatial derivatives on the magnetic
field. Hence, in order to obtain effects of comparable amount, the
characteristic length scale $L$ over which the magnetic field varies
must be of the order of the Compton wavelength of the neutrino
$L\sim1/m_\nu =1.9\cdot 10^{-5}\text{cm}\, (1\text{eV}/m_\nu)$.
Therefore, we expect the above-presented examples to be of greater
importance.

Except for mass differences, the above-discussed examples are not
directly sensitive to different neutrino flavors, since $\gA^2=1/4$
holds for all flavors ($\gA=1/2$ for electron neutrinos and $\gA=-1/2$
for others). A different example can be inferred from the axial
current Eq. \re{5}: consider a spatially constant electromagnetic
vacuum field varying in time with non-vanishing $\mathbf{E\cdot B}\neq
0$. Its contribution to the axial charge density is $\frac{\gA}{e}
\langle j_5^0\rangle =\frac{\alpha}{6\pi}\frac{\gA}{m^2} \frac{d}{dt}
\mathbf{E\cdot B}$. Therefore, such a field configuration is in
complete analogy to a polarized medium. In this way, a neutrino
propagating in such a field can be subject to an enhancement of flavor
oscillations, similarly to a propagation in matter.

\section*{Acknowledgments}

We would like to thank Professor W. Dittrich for helpful discussions
and for carefully reading the manuscript. This work was supported by
Deutsche Forschungsgemeinschaft under DFG Di 200/5-1.

\end{document}